\documentclass[pra,superscriptaddress,twocolumn]{revtex4-2}

\usepackage{enumerate}
\usepackage{amsfonts,amssymb,amsmath}
\usepackage[]{graphics,graphicx,epsfig}
\usepackage{amsthm}
\usepackage{graphicx}
\usepackage{dcolumn}
\usepackage{natbib}
\usepackage{color}
\usepackage{multirow}
\usepackage{ulem}
\usepackage{bm}
\usepackage{url}

\usepackage[colorlinks = true, citecolor=blue, linkcolor=blue, urlcolor=blue]{hyperref}
\newcommand*{\fullref}[1]{\hyperref[{#1}]{\autoref*{#1} \nameref*{#1}}}

\begin{document}

%%%%%%%%%%%%%%%%%%%%%%%%%%%%%%%%%%%%%%%%%%%%%%%%%%%%%%%%%

\title{Non-classicality Primitive in a Quasi-probabilistic Toy Model}
\author{Kelvin Onggadinata}
\affiliation{Centre for Quantum Technologies,
National University of Singapore, 3 Science Drive 2, 117543 Singapore,
Singapore}
\affiliation{Department of Physics,
National University of Singapore, 3 Science Drive 2, 117543 Singapore,
Singapore}

\author{Pawe{\l} Kurzy{\'n}ski}
\email{pawel.kurzynski@amu.edu.pl}
\affiliation{ Institute of Spintronics and Quantum Information, Faculty of Physics, Adam Mickiewicz University, Uniwersytetu Pozna{\'n}skiego 2, 61-614 Pozna\'n, Poland}
\affiliation{Centre for Quantum Technologies,
National University of Singapore, 3 Science Drive 2, 117543 Singapore,
Singapore}

\author{Dagomir Kaszlikowski}
\email{phykd@nus.edu.sg}
\affiliation{Centre for Quantum Technologies,
National University of Singapore, 3 Science Drive 2, 117543 Singapore,
Singapore}
\affiliation{Department of Physics,
National University of Singapore, 3 Science Drive 2, 117543 Singapore,
Singapore}

\date{\today}

%%%%%%%%%%%%%%%%%%%%%%%%%%%%%%%%%%%%%%%%%%%%%%%%%%%%%%%%%%%%%%%%%%%%%%%%%%%%%%%%%%%%%%%%%%%%%%%%%%%%%%%%%%%%%%%%

\begin{abstract}
 
We demonstrate a basic non-classical effect in a quasi-probabilistic toy model with local Alice and Bob who share classical randomness. Our scenario differs from the orthodox demonstrations of non-classicality such as violations of Bell inequalities where both local observers have a free will and randomly choose their measurement settings. The core of the argument are modified algorithms by Abramsky and Brandenburger [in {\it Horizons of the Mind}, Springer, Cham (2014)], and Pashayan {\it et. al.} [Phys. Rev. Lett. 115, 070501 (2015)] we use to show that if Bob deterministically performs a quasi-stochastic operation, Alice and Bob require classical communication to simulate it. 
\end{abstract}

\maketitle

%%%%%%%%%%%%%%%%%%%%%%%%%%%%%%%%%%%%%%%%%%%%%%%%%%%%%%%%%%%%%%%%%%%%%%%%%%%%%%%%%%%%%%%%%%%%%%%%%%%%%%%%%%%%%%%%

\section{Introduction}

In recent years, an operational approach to physics has yielded a multitude of intriguing results \cite{barrett2007information, chiribella2010probabilistic,fritz2012beyond}. This approach centers on what an experimenter can achieve with a system of interest. It dissects each experiment into three fundamental stages:
\begin{enumerate}
    \item Preparation: fixing a desired state $s$.
    \item Transformation: changing $s$ to $s'$.
    \item Measurement: information acquisition (partial or full) about $s'$.
\end{enumerate}
One of the primary objectives of this approach is to investigate how fundamental principles imposed on both the system and the experimenter influence these three stages. This approach works for a wide range of physical theories, including quantum and classical physics. A particularly interesting example of this approach is the Generalized Probabilistic Theory (GPT) \cite{plavala2023general}. In this framework, states are probability distributions of potential outcomes of measurable properties. These distributions can undergo transformations into other distributions, followed by measurements.

Here we study a simple instance of a GPT where the system's state is characterized by a non-negative probability distribution of three states of the system $[0,1,2]$, i.e., $s\equiv (p_0,p_1,p_2)$, $1\geq p_i\geq 0$. What sets our model apart from classical systems are state transformations $S$ in stage (2). We assume they are quasi-stochastic processes, i.e., stochastic processes with negative probabilities of transition between the states $i$ and $j$, $S_{i,j}$ ($i,j=0,1,2$ and $\sum_i S_{i,j}=1$) where at least one of the probabilities $S_{i,j}$ is negative.  We emphasize that states $s$ and $s'$ in the stages (1) and (3) are non-negative probability distributions because, unlike transformations $S$, $s$ and $s'$ can be directly measured. However exotic it may appear, quantum theory can be framed in this manner thanks to Wigner \cite{wigner1932quantum} whose ideas were later generalized to so-called {\it frames} \cite{ferrie2008frame, ferrie2009framed}. As a side remark, note that in the full-fledged frame formalism, states $s$ can be represented by negative probabilities too.

\begin{figure}[t]
    \centering
    \includegraphics[width=1.0\linewidth]{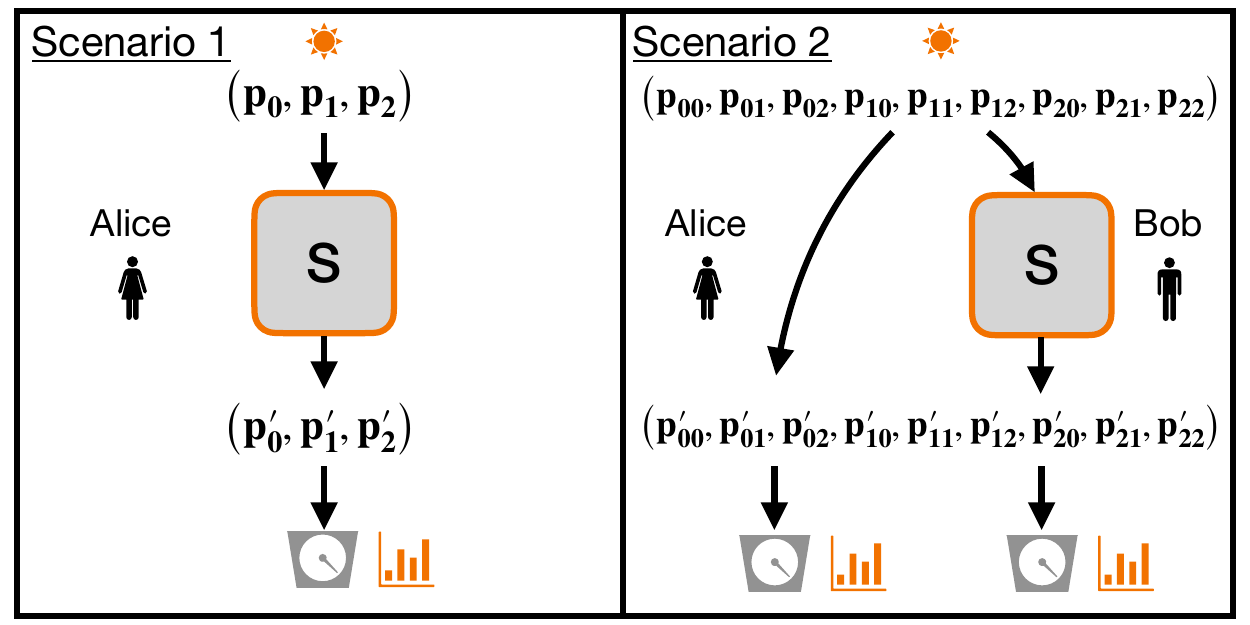}
    \caption{Schematic representation of the two scenarios under investigation in this study. Scenario 1: a source prepares a three-level system based on the probability distribution $(p_0, p_1, p_2)$. The distribution undergoes a quasiprobabilistic operation denoted as $S$, resulting in the transformed distribution $(p'_0, p'_1, p'_2)$. Subsequently, the system is subjected to measurement. Scenario 2: A source prepares two three-level systems in a correlated state described by the distribution $(p_{00}, p_{01}, p_{02}, p_{10}, p_{11}, p_{12}, p_{20}, p_{21}, p_{22})$. One of these systems is sent to Alice, while the other is sent to Bob. Bob performs the quasiprobabilistic operation $S$ on his system, leading to a transformed distribution $(p'_{00}, p'_{01}, p'_{02}, p'_{10}, p'_{11}, p'_{12}, p'_{20}, p'_{21}, p'_{22})$. Both systems are subsequently subjected to measurement. \label{f1}}
\end{figure}

We ask a fundamental question for this setup: Can the quasi-stochastic transformation $S$ be classically simulated? To this end, we investigate two distinct scenarios, as illustrated in Fig. \ref{f1}. In scenario 1, a single experimenter, Alice, has access to a source of a single three-level system. The source initially prepares the system in a state described by a distribution $(p_0, p_1, p_2)$. Subsequently, this state undergoes a quasi-probabilistic operation $S$. The system's state $(p_0,p_1,p_2)$ is transformed to a new distribution $(p'_0, p'_1, p'_2)$. We remind the reader that these are non-negative probability distributions. Finally, Alice conducts a measurement and records one of three possible results: 0, 1, or 2. In scenario 2, two experimenters, Alice and Bob, share a pair of three-level systems initially prepared in a correlated state, as described by a non-negative distribution $(p_{00}, p_{01}, p_{02}, p_{10}, p_{11}, p_{12}, p_{20}, p_{21}, p_{22})$. However, in this case, only Bob applies the quasi-probabilistic transformation $S$ to his part of the system. This operation leads to a final non-negative distribution $(p'_{00}, p'_{01}, p'_{02}, p'_{10}, p'_{11}, p'_{12}, p'_{20}, p'_{21}, p'_{22})$. After Bob's transformation, both Alice and Bob perform measurements on their respective systems.

We show a method to locally simulate the transformation $S$ in scenario 1 using classical stochastic transformations and post-selection. However, when dealing with spatially separated systems in scenario 2, such a simulation is not possible. This finding exposes a fundamental non-classicality of quasiprobabilistic models.

%%%%%%%%%%%%%%%%%%%%%%%%%%%%%%%%%%%%%%%%%%%%%%%%%%%%%%%%%%%%%%%%%%%%%%%%%%%%%%%%%%%%%%%%%%%%%%%%%%%%%%%%%%%%%%%%

\section{Quasiprobabilities}

Quasi-probabilities are inevitable when one wants to get rid of quantum probability amplitudes \cite{wigner1932quantum}, i.e., when one forces a classical description on quantum systems. One of the most widely recognized example of quasi-probabilities in quantum theory is the Wigner function \cite{wigner1932quantum}, which represents a quasi-probability distribution over classical phase space. A more contemporary application of quasi-probabilities can be found in studies on nonlocality and contextuality \cite{booth2022contextuality,morris2022witnessing,onggadinata2023simulations,onggadinata2023reexamination}. In these investigations, violations of corresponding Bell-type inequalities imply the negativity of joint probability distributions involving all observables considered in specific scenarios. Furthermore, recent research indicates that quasi-probabilities can offer computational speedup \cite{howard2014contextuality}.

A shared characteristic of quasi-probabilistic models is that negative probabilities are not directly observable during the measurement stage. This characteristic frees us from engaging in ontological debates regarding their interpretation. It's worth noting that a similar ontological challenge arises in the standard Hilbert space formulation of quantum theory — the meaning of the wave function remains controversial, a fact evident from numerous interpretations of quantum theory. However, a pragmatic resolution is to use wave function as a mathematical tool to calculate probabilities of experimental events. Frames formalism is yet another mathematical tool.

In this work, we examine a system described by a single probability distribution and introduce its quasi-probabilistic transformation. As demonstrated in our previous work \cite{onggadinata2023qubits}, the simplest nontrivial and consistent quasi-probabilistic dynamics can only be identified for a system with three or more possible states. Therefore, we begin by considering a single three-level system, with levels labeled as 0, 1, and 2. Its state is represented by a probability vector
\begin{equation}
    {\mathbf{p}}=\begin{pmatrix}
        p_0 \\ p_1 \\ p_2
    \end{pmatrix} \geq 0,
\end{equation}
where $p_0+p_1+p_2=1$.

Next, we introduce the following transformation matrix
\begin{equation}
    \mathbf{S}=\frac{1}{3}\begin{pmatrix} 2 & -1 & 2 \\ 2 & 2 & -1 \\ -1 & 2 & 2 \end{pmatrix},
\end{equation}
where $S_{i,j}$ $(i,j=0,1,2)$ is the quasi-probability of transition from the state $j$ to $i$. For example, the quasi-probability of transition from 0 to 1 is $2/3$ and that of transition from 1 to 0 is $-1/3$ etc. Note that $\mathbf{S}$ is quasi-bistochastic, since each of its rows and columns sums to one. It is also a reversible matrix that is quasi-bistochastic as well.  

\begin{figure}[t]
    \centering
    \includegraphics[width=0.8\linewidth]{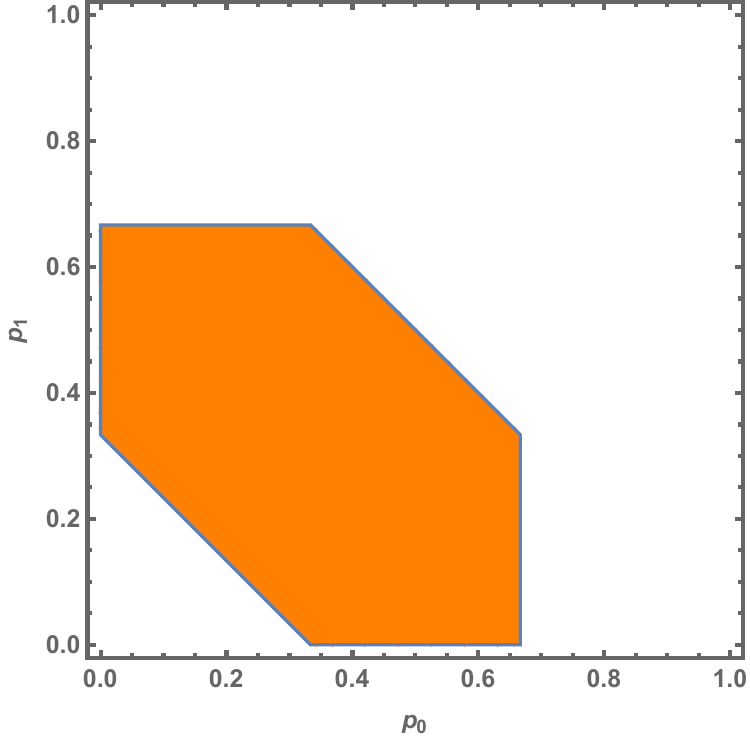}
    \caption{The representation of the region of probabilities that remain nonnegative under the action of $\mathbf{S}$ in the $p_0p_1$-plane. \label{f2}}
\end{figure}

Let us define
\begin{equation}\label{S}
   \mathbf{S} \cdot  {\mathbf{p}} = {\mathbf{p}}'=\begin{pmatrix}
        p'_0 \\ p'_1 \\ p'_2
    \end{pmatrix}.
\end{equation}
Quasi-bistochasticity implies that $p'_0 + p'_1 + p'_2 = 1$, but it does not guarantee that $\mathbf{p}' \geq 0$. For instance, if $\mathbf{p} = (1, 0, 0)^T$, then $\mathbf{p}' = \frac{1}{3}(2, 2, -1)^T$. To address this, we define a subset of probability vectors that remain non-negative under the action of $\mathbf{S}$. Notably, we observe that $\mathbf{S}^{n+6} = \mathbf{S}^n$, indicating that $\mathbf{S}$ is periodic with a period of six, and $\mathbf{S}^6 = \openone$. Applying a straightforward positivity condition allows us to represent the subset of allowed probabilities as a convex region in the $p_0p_1$-plane (see Fig. \ref{f2}). The distributions within this region can be expressed as convex combinations of the following extreme points
\begin{eqnarray}
    & &\mathbf{e}_0 = \frac{1}{3}\begin{pmatrix}
        2 \\ 1 \\ 0
    \end{pmatrix},~~~~\mathbf{e}_1 = \frac{1}{3}\begin{pmatrix}
        1 \\ 2 \\ 0
    \end{pmatrix},~~~~\mathbf{e}_2 = \frac{1}{3}\begin{pmatrix}
        0 \\ 2 \\ 1
    \end{pmatrix}, \nonumber \\
    & &\mathbf{e}_3 = \frac{1}{3}\begin{pmatrix}
        0 \\ 1 \\ 2
    \end{pmatrix},~~~~\mathbf{e}_4 = \frac{1}{3}\begin{pmatrix}
        1 \\ 0 \\ 2
    \end{pmatrix},~~~~\mathbf{e}_5 = \frac{1}{3}\begin{pmatrix}
        2 \\ 0 \\ 1
    \end{pmatrix}, \label{extr}
\end{eqnarray}
where 
\begin{equation}\label{transf}
\mathbf{e}_{i+1}=\mathbf{S}\cdot\mathbf{e}_i,
\end{equation}
and $\mathbf{e}_{6}\equiv \mathbf{e}_{0}$.

We now have a set of permissible states that remain non-negative under the action of $\mathbf{S}$. From the standpoint of the operational framework, we have a classical probabilistic description of the system in stages (1) and (3). We can now treat $\mathbf{S}$ as a black box, the inner workings of which lie beyond our scrutiny (similarly, you cannot look `inside' of a quantum interferometer without destroying interference). Nevertheless, a valid question arises: Can this black box be effectively simulated using classical methods? If yes, the entire operational model is classical. Else, $\mathbf{S}$ has {\it something} non-classical about it and we would like to understand and quantify it. 

%%%%%%%%%%%%%%%%%%%%%%%%%%%%%%%%%%%%%%%%%%%%%%%%%%%%%%%%%%%%%%%%%%%%%%%%%%%%%%%%%%%%%%%%%%%%%%%%%%%%%%%%%%%%%%%%

\section{Simulation}

%%%%%%%%%%%%%%%%%%%%%%%%%%%%%%%%%%%%%%%%%%%%%

\subsection{Scenario 1}

It is enough to focus on the states $\mathbf{e}_0$, $\mathbf{e}_2$, $\mathbf{e}_4$ to show that $\mathbf{S}$ cannot be simulated by a non-negative stochastic operation
\begin{equation}
\mathbf{\tilde{S}}=\begin{pmatrix} p & p' & p'' \\ q & q' & q'' \\ 1-p-q & 1-p'-q' & 1- p''-q'' \end{pmatrix}.
\end{equation}
Note that Eq. (\ref{transf}) and simulability of $\mathbf{S}$ imply $\mathbf{\tilde{S}}\cdot \mathbf{e}_0=\mathbf{e}_1$, $\mathbf{\tilde{S}}\cdot \mathbf{e}_2=\mathbf{e}_3$, and $\mathbf{\tilde{S}}\cdot \mathbf{e}_4=\mathbf{e}_5$. This, together with non-negativity of $\mathbf{\tilde{S}}$, gives us: 
\begin{equation}
1-p-q=1-p'-q'=p'=p''=q=q''=0.
\end{equation}
As a result
\begin{equation}
p=q'=1-p''-q''=1,
\end{equation}
implying that $\mathbf{\tilde{S}}$ is the identity matrix. This is evidently incorrect. End of the proof.

Since $\mathbf{S}$ cannot be simulated by a non-negative stochastic process, which is a linear transformation, there is still a chance that it can be simulated by some nonlinear probabilistic operation. However, an implementation of nonlinear probabilistic operations require manipulations on multiple copies of the system. For example, one can consider a state-dependent transformation 
\begin{equation}
\mathbf{\tilde{S}}(\mathbf{p})\cdot \mathbf{p} = \mathbf{p}',
\end{equation}
such that
\begin{eqnarray}
& & \mathbf{\tilde{S}}(\mathbf{e}_0) = \begin{pmatrix} 0 & 1 & 0 \\ 1 & 0 & 0  \\ 0 & 0 & 1 \end{pmatrix},~~\mathbf{\tilde{S}}(\mathbf{e}_2)=\begin{pmatrix} 1 & 0 & 0 \\ 0 & 0 & 1  \\ 0 & 1 & 0 \end{pmatrix}, \nonumber \\ & & \mathbf{\tilde{S}}(\mathbf{e}_4)=\begin{pmatrix} 0 & 0 & 1 \\ 0 & 1 & 0  \\ 1 & 0 & 0 \end{pmatrix}.
\end{eqnarray}
However, since $\mathbf{e}_0$, $\mathbf{e}_2$ and $\mathbf{e}_4$ cannot be perfectly distinguished, one has to perform measurements on multiple copies to estimate the input state. 

The above method is input-dependent and in case of an arbitrary input state the number of distinct operations $\mathbf{\tilde{S}}(\mathbf{p})$ may be infinite. Therefore, we propose an input-independent method to simulate quasiprobabilistic processes. It is based on the idea by Abramsky and Brandenburger \cite{abramsky2014operational}, who proposed that negative probabilities can be simulated with classical probabilities by adding an additional two-level system that labels which events are assigned negative and which non-negative probabilities. We stress that our method works for arbitrary quasi-stochastic processes that acts on an arbitrary d-level system. 

First, observe that any quasi-stochastic matrix $\mathbf{S}$ can be split into positive and negative parts with the help of so-called {\it nebit} \cite{kaszlikowski2021little}, the simplest quasi-probabilistic system with binary outcome probability distribution $(q^+,-q^-)$:
\begin{equation}\label{split}
\mathbf{S} =  q^+ \mathbf{S}^{+} - q^- \mathbf{S}^{-}, 
\end{equation}
where $ \mathbf{S}^{\pm}$ are some (non-negative) stochastic matrices, $q^{\pm} \geq 0$ and $q^+ - q^- = 1$. For an arbitrary input state $\mathbf{p}$ we define $\mathbf{p}^+ =\mathbf{S}^+ \cdot \mathbf{p}$ and $\mathbf{p}^- =\mathbf{S}^- \cdot \mathbf{p}$. We get
\begin{equation}\label{split2}
\mathbf{S} \cdot \mathbf{p} = q^+ \mathbf{p}^+ - q^- \mathbf{p}^- = \mathbf{p}'.
\end{equation}
This looks like applying $ \mathbf{S}^{+}$ with inflated nebit probability $q^+$ and $ \mathbf{S}^{-}$ with negative nebit probability $-q^-$. 

The above splitting is non-unique. One possibility to find it is to look for a (non-negative) stochastic matrix $ \mathbf{S}^{-}$  and a positive constant $\alpha$ for which
\begin{equation}
\mathbf{S'}^+ = \mathbf{S} + \alpha  \mathbf{S}^{-}  \geq 0. 
\end{equation}
Because each column of $\mathbf{S}$ and $\mathbf{S}^-$ sums to one, each column of $\mathbf{S'}^+$ sums to $1+\alpha$. Therefore, $\mathbf{S}^+ = \frac{1}{1+\alpha} \mathbf{S'}^+$ is a stochastic matrix too. Finally, we arrive at Eq. (\ref{split}) if we define  $q^+ = 1+\alpha$ and $q^- = \alpha$. 

The simulation works as follows. We define
\begin{equation}
r = \frac{q^+}{q^+ + q^-},
\end{equation}
and introduce a single bit that is zero with probability $r$ and one with probability $1-r$. Note that $r>1-r$, since $q^+ - q^-=1$. If this bit is zero we apply $\mathbf{S}^+$ to the input and if it is one we apply $\mathbf{S}^-$. This can be written as
\begin{eqnarray}
& &\left( \begin{pmatrix} 1 & 0 \\ 0 & 0 \end{pmatrix} \otimes \mathbf{S}^+  +  \begin{pmatrix} 0 & 0 \\ 0 & 1 \end{pmatrix} \otimes \mathbf{S}^-  \right) \cdot  \begin{pmatrix} r \\ 1-r \end{pmatrix} \otimes  \mathbf{p} = \nonumber \\
& &  \begin{pmatrix} r \\ 0 \end{pmatrix} \otimes  \mathbf{p}^+ + \begin{pmatrix} 0 \\ 1-r \end{pmatrix} \otimes  \mathbf{p}^- =  \begin{pmatrix} r \mathbf{p}^+ \\ (1-r)\mathbf{p}^- \end{pmatrix}.
\end{eqnarray}  

After the above controlled operation we perform a measurement and register the outcome of the bit $b=0,1$ and the system $x=0,1,\ldots, d-1$. We repeat this procedure multiple times and collect measured outcomes in a table. The next step of the simulation is post-selection -- we remove some data from the table. The removal procedure is based on the following rule. For each measurement event $\{b=1,x\}$ we look for an event $\{b=0,x\}$ and remove both from the table. The schematic representation of the removal procedure is shown below.
\begin{center}
\begin{tabular}{ | c | c | } 
 \hline
   b & x \\ 
  \hline
  \hline
   0 & 2  \\ 
  \hline
   0 & 0  \\
  \hline
   0 & 1  \\
  \hline
   1 & 2  \\
  \hline
  \vdots & \vdots   \\
  \hline
\end{tabular} ~~ $\rightarrow$ ~~ 
\begin{tabular}{ | c | c | } 
 \hline
   b & x \\ 
  \hline
  \hline
   {\color{red} 0} &  {\color{red}  2}  \\ 
  \hline
   0 & 0  \\
  \hline
   0 & 1  \\
  \hline
  {\color{red}  1} &  {\color{red} 2}  \\
  \hline
  \vdots & \vdots   \\
  \hline
\end{tabular} ~~ $\rightarrow$ ~~ 
\begin{tabular}{ | c | c | } 
 \hline
   b & x \\ 
  \hline
   \hline
   0 & 0  \\
  \hline
   0 & 1  \\
  \hline
  \vdots & \vdots   \\
  \hline
\end{tabular} 
\end{center}
The goal is to remove from the table all events for which $b=1$. If we succeed, we use the final table to evaluate the probabilities $p'_x$. If we fail, i.e., there are still some events $\{b=1,x\}$ in the table for which there is no pair $\{b=0,x\}$, we abandon the simulation.

Now, we prove that the obtained probabilities $p'_x$ approximate the ones we would obtain if we applied the quasi-stochastic operation $\mathbf{S}$. First, note that due to Eq. (\ref{split2}) the action of  $\mathbf{S}$ gives $p'_x=q^+ p^+_x - q^- p^-_x$. On the other hand, for sufficiently large number of experiments $N$, the number of events $\{b=0,x\}$ in the initial table is $N_{\{b=0,x\}} \approx N r p^+_x$ and the number of events $\{b=1,x\}$ is $N_{\{b=1,x\}}\approx N (1-r) p^-_x$. However, after the removal procedure the number of remaining events $\{b=0,x\}$ is
\begin{eqnarray}
& &N_{\{b=0,x\}}-N_{\{b=1,x\}} \approx N \left(rp^+_x - (1-r)p^-_x\right)=  \nonumber \\
& & \frac{N}{q^+ + q^-}\left(q^+ p^+_x - q^- p^-_x\right) \approx N'_{\{b=0,x\}}.
\end{eqnarray}
In addition, note that the total number of events for which $b=1$ is approximately $N(1-r)$, therefore after the removal the total number of events in the table changes from $N$ to
\begin{equation}
N' \approx N-2N(1-r) = N\left(\frac{q^+ - q^-}{q^+ + q^-}\right) = \frac{N}{q^+ + q^-}.
\end{equation}
As a result,
\begin{equation}
p'_x \approx \frac{N'_{\{b=0,x\}}}{N'}=q^+ p^+_x - q^- p^-_x.
\end{equation}

Let us also comment on how to interpret a possible failure of our simulation protocol. This happens if the table contains more events $\{b=1,x\}$ than events $\{b=0,x\}$, i.e.,  $N_{\{b=0,x\}} < N_{\{b=1,x\}}$, which gives
\begin{equation}
q^+ p^+_x  < q^- p^-_x. 
\end{equation}
However, this means that the action of $\mathbf{S}$ would generate observable negative probability, since in this case $p'_x<0$. Therefore, failure of our protocol prevents such possibilities. Note, that our initial assumption that $\mathbf{p}$ is chosen from a set that remains non-negative under the action of $\mathbf{S}$ guarantees that for sufficiently large number of experiments our protocol should succeed with probability close to one. 

Finally, we can come back to our original operation $\mathbf{S}$ from Eq. (\ref{S}). To simulate it with our protocol we can choose
\begin{equation}
\mathbf{S} = \frac{4}{3} \begin{pmatrix} 1/2 & 0 & 1/2 \\ 1/2 & 1/2 & 0  \\ 0 & 1/2 & 1/2 \end{pmatrix} - \frac{1}{3}  \begin{pmatrix} 0 & 1 & 0 \\ 0 & 0 & 1  \\ 1 & 0 & 0 \end{pmatrix} =  \frac{4}{3} \mathbf{S}^{+} - \frac{1}{3} \mathbf{S}^{-}.
\end{equation}
This choice leads to $r=4/5$.

%%%%%%%%%%%%%%%%%%%%%%%%%%%%%%%%%%%%%%%%%%%%%

\subsection{Scenario 2}

Consider two correlated copies of the previous three-level system shared between Alice and Bob and prepared in a state
\begin{equation}
\mathbf{p}_{AB} = \frac{1}{3} \begin{pmatrix} 1 \\ 0 \\ 0  \end{pmatrix} \otimes \mathbf{e}_0 + \frac{1}{3} \begin{pmatrix} 0 \\ 1 \\ 0  \end{pmatrix} \otimes \mathbf{e}_2 + \frac{1}{3} \begin{pmatrix} 0 \\ 0 \\ 1  \end{pmatrix} \otimes \mathbf{e}_4.
\end{equation}
This non-negative probability distribution says that if Alice registers $0$ then Bob's system is in the state $\mathbf{e}_0$, if she registers $1$ Bob's state is $\mathbf{e}_2$, and if she registers $2$ Bob's state is $\mathbf{e}_4$. 

Interestingly, $\mathbf{p}_{AB}$ cannot be represented as a convex combination of products of the extreme states 
\begin{equation}
\mathbf{p}_{AB} \neq \sum_{i,j=1}^6 p_{i,j}   \mathbf{e}_i \otimes  \mathbf{e}_j,
\end{equation}
where all $p_{i,j}\geq 0$. This is because $\mathbf{p}_{AB}$ is a 9-dimensional probability vector with zeros at positions 3, 4 and 8. On the other hand none of the products $\mathbf{e}_i \otimes  \mathbf{e}_j$ has zeros at all of these three positions.

Nevertheless, it is easy to verify that $\mathbf{p}_{AB}$ remains non-negative under the local actions of  $\mathbf{S}$ from Eq. (\ref{S})
\begin{equation}
\forall_{n,m} ~~~~\mathbf{S}^n \otimes \mathbf{S}^m \cdot \mathbf{p}_{AB} \geq 0.
\end{equation}
It is evident now that $\mathbf{p}_{AB}$ has all the characteristic features of an entangled state. It may come as surprise to some that entanglement is represented by a non-negative probability distribution but this also happens in some frames representations of quantum theory itself (however, our model does not represent a quantum system).  

Now consider a transformed state  $\mathbf{p}_{AB}'=\openone \otimes \mathbf{S} \cdot \mathbf{p}_{AB} $, where  
\begin{equation}
\mathbf{p}_{AB}' = \frac{1}{3} \begin{pmatrix} 1 \\ 0 \\ 0  \end{pmatrix} \otimes \mathbf{e}_1 + \frac{1}{3} \begin{pmatrix} 0 \\ 1 \\ 0  \end{pmatrix} \otimes \mathbf{e}_3 + \frac{1}{3} \begin{pmatrix} 0 \\ 0 \\ 1  \end{pmatrix} \otimes \mathbf{e}_5.
\end{equation}
Can this transformation be simulated locally by Bob with classical methods?

Although we already know that Bob can simulate locally individual transformations $\mathbf{S}\cdot \mathbf{e}_i = \mathbf{e}_{i+1}$, here such simulation needs to take into account three different transformations at once.  Moreover, Bob's local state before and after transformation is
 \begin{equation}
\mathbf{p}_B=\mathbf{p}'_B= \frac{1}{3}( \mathbf{e}_0 +  \mathbf{e}_2 +  \mathbf{e}_4) = \frac{1}{3}( \mathbf{e}_1 +  \mathbf{e}_3 +  \mathbf{e}_5) = \frac{1}{3} \begin{pmatrix} 1 \\ 1 \\ 1  \end{pmatrix} ,
\end{equation}
therefore no local input-dependent nonlinear transformation $\mathbf{S}(\mathbf{p})$ can be used for simulation. This is also a reason why the above simulation of quasi-stochastic transformations with additional bit will not work. More precisely, in order to transform a distribution $\mathbf{p}_{AB}$ into $\mathbf{p}_{AB}'$, Bob needs to known Alice's statistics. The removal procedure needs to be applied to a table containing joint outcomes of both, Alice and Bob. We discuss this problem below.

Let us consider a previous simulation and assume that Alice also measures her state. The table of outcomes generated by such procedure may take the following form
\begin{center}
\begin{tabular}{ | c | c |c| } 
 \hline
   b & x & y\\ 
  \hline
  \hline
   0 & 2 & 1 \\ 
  \hline
   0 & 0 & 0 \\
  \hline
   0 & 1 & 2 \\
  \hline
   1 & 2 & 2 \\
  \hline
  \vdots & \vdots & \vdots   \\
  \hline
\end{tabular}
\end{center}
where the additional column $y$ corresponds to Alice's outcomes. If Bob has no access to $y$, his removal procedure may lead to
\begin{center}
\begin{tabular}{ | c | c |c| } 
 \hline
   b & x & y\\ 
  \hline
  \hline
   {\color{red}0} & {\color{red}2} & {\color{red}1} \\ 
  \hline
   0 & 0 & 0 \\
  \hline
   0 & 1 & 2 \\
  \hline
   {\color{red}1} & {\color{red}2} & {\color{red}2} \\
  \hline
  \vdots & \vdots & \vdots   \\
  \hline
\end{tabular} ~~ $\rightarrow$ ~~ 
\begin{tabular}{ | c | c | c | } 
 \hline
   b & x & y \\ 
  \hline
   \hline
   0 & 0 & 0 \\
  \hline
   0 & 1 & 2 \\
  \hline
  \vdots & \vdots & \vdots  \\
  \hline
\end{tabular} 
\end{center}
Simply speaking, Bob may pair and remove events that are globally different, which alters the final statistics. As a result, the final distribution will differ from $\mathbf{p}'$. 

The main observation is that in order to properly implement a simulation of a local quasi-stochastic operation on a spatially distributed correlated system a communication between the parties is necessary. More precisely, in our case Alice needs to send her outcomes to Bob. This is a fundamental non-local feature of quasi-probabilistic models unless there is a different classical simulation of quasi-probabilities, which is an open problem. 

%%%%%%%%%%%%%%%%%%%%%%%%%%%%%%%%%%%%%%%%%

\section{Conclusions}

We study a system with states $s$ represented as three-point, non-negative probability distributions ${\bf p}=(p_0,p_1,p_2)^T\geq 0$ and a discrete set of quasi-bistochastic transformations ${\bf S}^k$ ($k=0,1,2,\dots,5$, ${\bf S}^k$ is the $k$th power of matrix ${\bf S}$) that preserve non-negativity, i.e., ${\bf S}^k{\bf p}\geq 0$. This system does not have a quantum mechanical equivalent. 

First, we show how to classically simulate ${\bf S}^k$, using a modified algorithm presented in \cite{abramsky2014operational, pashayan2015estimating}. This algorithm requires post-selection. Next, we examine two local observers, Alice and Bob, sharing classical randomness via a bipartite, non-negative probability distribution ${\bf p}_{AB}$ and show that one cannot simulate Bob's local deterministically chosen operations $I\otimes {\bf S}^k$ without classical communication between Alice and Bob. 

This basic non-classical effect is a consequence of quasi-bistochasticity of Bob's operations. It remains an open question if the only known simulation algorithms in \cite{abramsky2014operational, pashayan2015estimating} can be improved to remove this non-classical effect.

%%%%%%%%%%%%%%%%%%%%%%%%%%%%%%%%%%%%%%%%%

\section*{Acknowledgements}

This research is supported by the National Research Foundation, Singapore, and A*STAR under the CQT Bridging Grant. PK is supported by the Polish National Science Centre (NCN) under the Maestro Grant no. DEC-2019/34/A/ST2/00081. We would like to thank A. Grudka, M. Karczewski, J. Stempin, A. W\'ojcik and J. W\'ojcik for stimulating discussions.

%%%%%%%%%%%%%%%%%%%%%%%%%%%%%%

\bibliographystyle{apsrev4-2}
\bibliography{ref.bib}

%%%%%%%%%%%%%%%%%%%%%%%%%%%%

\end{document}